\documentclass[11pt, a4paper]{article}
\usepackage{moriond,epsfig}
\usepackage{amsmath,bm}

\def\be{\begin{equation}}
\def\ee{\end{equation}}
\def\bea{\begin{eqnarray}}
\def\eea{\end{eqnarray}}

\def\lesssim{\ \hbox{\raise 2pt \hbox{$<$} \kern -13pt
                     \lower 3pt \hbox{$\sim$}}\ }
\def\greatersim{\ \hbox{\raise 2pt \hbox{$>$} \kern -13pt
                     \lower 3pt \hbox{$\sim$}}\ }
\begin{document}
\begin{flushright}
CERN-PH-TH/2007-071\\
\end{flushright}
\vspace*{1cm}
\title{Rescattering effects and the\\
 determination of the gluon density for x $\ll$ 1~\footnote{Talk 
 presented   at the XLII Rencontres de Moriond  
 (La Thuile,  March 2007).}}

\author{F.~Hautmann}
\address{CERN, PH-TH Division, 
Geneva, Switzerland and \\
Institut f{\" u}r Theoretische Physik,
Universit{\" a}t Regensburg, Germany}

\maketitle\abstracts{We consider 
the possible role of rescattering effects  in the 
determination of the gluon density for the LHC  from DIS data.  
We discuss a method that uses    results of s-channel 
calculations to  estimate  these effects, and   comment on 
potential applications to diffractive  and  
 multi-parton interactions. } %
\noindent

The Large Hadron Collider  will operate with very high gluon 
luminosities.    Production processes initiated by gluons contribute a great  many 
  events to a number of  cross sections  
of primary interest for the LHC physics program. 
Reliable predictions for these cross sections  depend on 
the determination of the gluon density in the 
proton and its accuracy.~\cite{thuncert,durham07,durham04,cteq06,alekhin05}    
As parton luminosities  rise steeply  for 
decreasing momentum fraction $x$, 
 a large number of events sample  the gluon density at  $x \ll 1$.  

It has long been known that 
the theoretical accuracy of gluon-density determinations  for 
 $x \ll 1$  is affected by 
 higher-loop $\ln (1/x)$ corrections to 
QCD evolution equations. See e.g. Ref.~\cite{ehw} 
for  early numerical investigations.   
 The study of these corrections motivates   current 
 searches for   
 evolution schemes 
 (see Refs.~\cite{thorne06,abf06,ccss06}, and references 
therein) 
that incorporate the resummation of  $\ln (1/x)$ contributions 
at the next-to-leading-logarithmic  accuracy.~\cite{CH94,flcc}  
An improved theoretical control     on the $x \ll 1$  
region is expected  from the   inclusion of  these terms.

Because the DIS data used at present to extract 
the gluon density  for $x < 10^{-2}$  do not have very high $Q^2$, 
 it is natural to ask whether  
 non-negligible effects on the theoretical accuracy 
 may also come from  corrections that are suppressed by powers of 
  $1 / Q^2$ but are potentially enhanced as $x \to 0$. 
 These could  affect the determination of the gluon  density  
  $f_g$  primarily  through a contribution $\delta$ to the 
  $Q^2$-derivative of the $F_2$ structure function, 
\begin{equation}
\label{df2}
{{d F_2} \over { d \ln Q^2 }} \simeq P_{q g} 
\otimes f_g \ \left[ 1 + \delta
\right] + \ {\rm{quark}} \; \ {\rm{term}} \; 
\hspace*{0.2 cm}  ,
\hspace*{0.2 cm} 
\delta \simeq \sum_{k \geq 1} \ a_k \ (\alpha_s  \ {1 \over x^\nu} \ 
{\Lambda^2 \over Q^2})^k \;\;  . 
\end{equation} 
Here $P_{q g} $ is the  perturbative  
gluon-to-quark 
evolution kernel, and the correction $\delta$  
arises  from multi-parton correlation terms in the operator product 
 expansion, 
\begin{equation}
\label{opefac} 
F_2 = C \otimes f + { 1 \over Q^2}  \ 
C^{(4)} \otimes f^{(4)} + \dots \;\; .    
\end{equation}
The enhanced $x \to 0$ behavior in Eq.~(\ref{df2}) 
can be produced from graphs with multiple gluon 
scatterings, and 
is consistent 
with observations of approximate geometric scaling 
in low-$x$ data~\cite{geoscal} and with models 
 for saturation.~\cite{golecrev,iancu06,mue99}

Standard methods to take account of multiple scatterings are 
s-channel methods (see e.g.  
 the lectures in Ref.~\cite{mue99}),  
essentially orthogonal  to  
those of the 
parton picture. The basic degrees of freedom in 
the s-channel picture are 
 described  by correlators of 
eikonal Wilson lines --- at the simplest level, two-point correlators, 
interpretable as color dipoles. 
To identify the correction  from rescattering graphs 
to the 
parton result  in Eq.~(\ref{df2}),   
 a sufficiently precise 
``dictionary"  is needed to connect the two pictures. 
Ref.~\cite{hs07} presents an  approach to 
analyze this connection. 

The method is  based on 
constructing explicitly  an s-channel 
representation for the renormalized parton distribution function 
   in terms of 
 Wilson-line matrix elements, convoluted with 
 lightcone wave functions.  In    this       
  representation  the quark distribution $f_q$ 
is given by  the coordinate-space convolution 
\begin{equation}
\label{convfq}
x f_q ( x , \mu) = \int {d{\bm z}} \int {d{\bm b}} \ 
u (\mu ,  {\bm z} ) \ \Xi( {\bm z}, {\bm b}) \;\; ,  
\end{equation}
where 
$\Xi$ is the hadronic  matrix element of  eikonal-line 
operators,  
\begin{equation}
\label{xidef}
\begin{split} 
\Xi( {\bm z}, {\bm b}) = 
\int [ d P^\prime ] \   
\langle P'|\frac{1}{N_c} \ {}& {\rm Tr}\{1 - 
V^{\dagger}( {\bm b} + {\bm z}/2)\,
V({\bm b}-{\bm z}/2)
\} |P \rangle \hspace*{0.2 cm} ,
\\&  V({\bm z}) = {\cal P}\exp\left\{
-ig\int_{-\infty}^{+\infty}dz^- { A}^+_a(0,z^-,{\bm z}) t_a
\right\} \;\;  ,     
\end{split} 
\end{equation}
${\bm z}$ is the transverse separation between the eikonal lines, 
${\bm b}$ is the impact parameter, 
and the  
function $u (\mu ,  {\bm z} )$ is evaluated explicitly 
 in Ref.~\cite{hs07} at one loop 
using the $\overline {\rm MS}$   scheme for 
the renormalization of the 
ultraviolet divergences ${\bm z} \to 0$. 

The  representation  (\ref{convfq}), once evaluated in a  well-prescribed 
renormalization scheme,  is the key ingredient that  allows one  to 
relate~\cite{hs07,plb06}  results of s-channel 
calculations for   structure functions  to the OPE 
factorization  (\ref{opefac}),   
and, in particular, to identify the power-suppressed 
corrections arising from  the 
multiple gluon scatterings, treated  in 
the high-energy approximation of Eq.~(\ref{xidef}). 
These corrections are found  to  
 depend on  moments  of $\Xi$, schematically  in the   form 
\begin{equation}
\label{schemat}
{{d F_2} \over { d \ln Q^2 }} =
\left( {{d F_2} \over { d \ln Q^2 }} \right)_{\rm{LP}}
+ \sum_{n=1}^\infty \ R_n \ {\lambda^2 (n) \over {( Q^2 )^n}} \; \; , 
\end{equation}
where the first term in the right hand side is the leading-power 
parton result, and the  $\lambda^2 $ in the subleading terms  
 are given  by the  analytically continued  moments 
\begin{equation}
\label{moments}
\lambda^2 (- v)   = {1 \over \Gamma ( v ) }
\int   {{d{\bm z}} \over {\pi {\bm z}^2}}
\ ({\bm z}^2 )^{v - 1} \
 \int d {\bm b} \ \Xi( {\bm z}, {\bm b}) \; \; . 
\end{equation}
The coefficients $R_n$ are evaluated to order $\alpha_s$, as functions of $x$ and 
$\ln Q^2$,  from the lightcone wave functions, while the 
moments $\lambda^2 (n)$ are dimensionful nonperturbative parameters, 
 to  be determined from comparison with experimental data. 

\begin{figure}[htb]
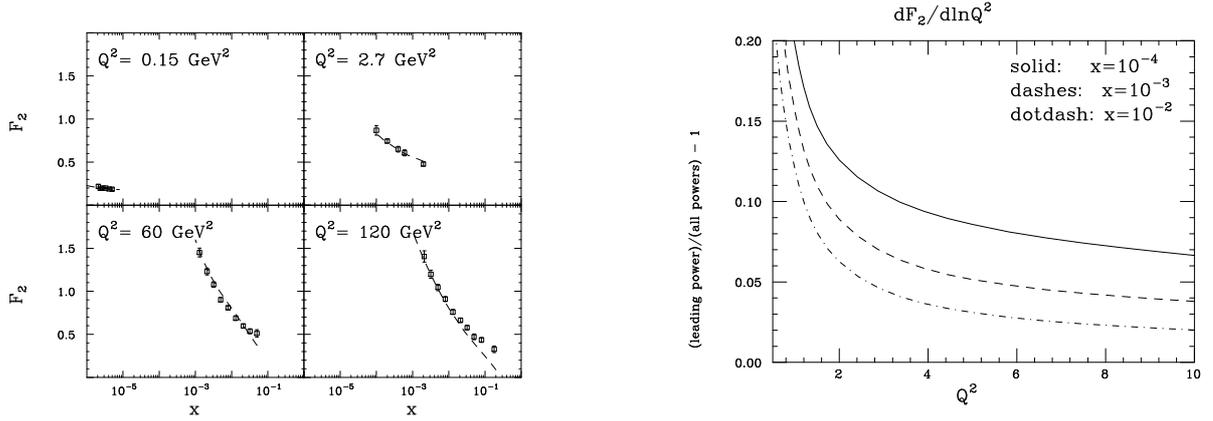

\vspace{55mm}
\includegraphics[scale=0.3,angle=90,bbllx=30,
bblly=400,bburx=-440,bbury=760]{dat_fourpan.ps}
\includegraphics[scale=0.32,angle=90,bbllx=70,
bblly=810,bburx=-400,bbury=1170]{allpow_lin.ps}
\caption{(left)  The result of fitting   the  $\lambda^2$  parameters 
to the  data~[18]; (right)  
power corrections to 
$d F_2 / d \ln Q^2$ versus $Q^2$ at different values of $x$. 
}
\label{fig:withtwo}
\end{figure}

In practice, the usefulness of the result in 
Eqs.~(\ref{schemat}),(\ref{moments}) comes from the fact 
that the hadronic matrix element  $\Xi$ can 
be related by a short-distance  expansion for 
${\bm z} \to 0$ to a well-prescribed integral of the 
gluon distribution function~\cite{hs07}. Then the  moments $\lambda^2$ 
can  be parameterized in terms of the factorization/renormalization scales 
at which the gluon distribution and the strong coupling are 
evaluated. 
These scales are to be 
 taken  of the order of the 
inverse mean transverse distance $1 / | {\bm z} |$, and can  be 
tuned to the data.  The result of 
 doing this  for   $F_2$ data~\cite{zdata} at  both low and high 
$Q^2$ is shown in  
Fig.~\ref{fig:withtwo} in the  left hand side plot.~\cite{plb06}  
 The corresponding 
 power correction  in Eq.~(\ref{schemat})  
is  plotted on the right hand side of   
Fig.~\ref{fig:withtwo}. Here we show 
the correction   
normalized to the full answer and multiplied by $-1$, using 
 CTEQ parton distributions.~\cite{cteq02}

Fig.~\ref{fig:withtwo} indicates that 
with physically natural choices of the parameters in the 
nonperturbative matrix elements (\ref{xidef})  one can achieve 
a sensible description of data for 
$x < 10^{-2}$  in a wide  range of $Q^2$  
 and  still have moderate power corrections to $d F_2 / d \ln Q^2$. Corrections 
turn out to be negative and 
below  20 $\%$ for $x \greatersim 10^{-4}$ and  $Q^2 \greatersim 1$ GeV$^2$. 
However, Fig.~\ref{fig:withtwo} 
also indicates that for small $x$ the corrections fall off slowly in the 
region of medium $Q^2$, $ Q^2 \simeq 1 - 10$ GeV$^2$, 
behaving effectively like  $1/Q^\lambda$ 
with
$   \lambda $ close to 1.~\cite{plb06} For 
instance, one has $ \lambda \simeq 1.2$ for the curve 
$x = 10^{-3}$ in the right-hand side plot of Fig.~\ref{fig:withtwo}.  
As a consequence of the slowly decreasing behavior, 
the power corrections  stay on the order of 
$ 10 \%$ up  to $Q^2$ of a few GeV$^2$ for $x 
 \lesssim 10^{-3}$. 
This  slow fall-off    differs 
from parameterizations of higher twist commonly   used 
in global analyses (see e.g. Ref.~\cite{durham04}), and may be relevant 
for phenomenology as it affects the medium $Q^2$ region of the   
data used  to extract $f_g$ at low $x$. 

Corrections larger than for $F_2$ are found for the longitudinal 
 component $F_L$.~\cite{plb06}  
This provides additional 
 motivation for the forthcoming $F_L$ measurements~\cite{hera-fl},   
as well as  fits~\cite{durham-fl}   
 investigating   
power-like   terms in $F_L$ (at both high and low x). 
We observe also that  Fig.~\ref{fig:withtwo}  is obtained using 
  NLO parton distributions,  and    
the decrease in the 
low-x gluon at NNLO~\cite{durham07} could be consistent with the 
possibility that NNLO parton distributions correspond to 
 smaller power corrections. However, the detailed interpretation of this 
behavior  will be subtle, as 
 distinctly different dynamics  drive the power-like and NNLO effects,   
 unlike the high-x case in the  analyses~\cite{durham04,gardirob}.

It is worth emphasizing that the  above results depend on the validity 
of the high-energy 
approximation and s-channel representation (\ref{convfq}),  and  
 the perturbation expansion for  $u (\mu ,  {\bm z} )$. 
The rationale for this expansion  
lies with the dynamical cut-off on large distances ${\bm z}$ 
imposed by unitarity requirements (``black disc" limit) on the 
correlator $\Xi$.~\cite{iancu06,mue99,hs07}
But the size of this cut-off at collider  energies is  difficult to determine. 
The highest sensitivity to it may  come from measurements of the diffractive part 
of the DIS cross section.~\cite{abra,h1diffr,h1diffjet} 
In this case the s-channel representation is bilinear in $\Xi$. 
The comparison~\cite{hs00}  of diffractive  data with theoretical predictions 
based on diffractive parton distributions  
indicates that the dynamical cut-off lies at  substantially  higher momenta  
for color-octet eikonal-line matrix elements than for color-triplet.  
 That is, gluons' shadow is stronger. 
See e.g. Ref.~\cite{kopel06} for a recent discussion.
In diffractive DIS 
this can be linked~\cite{abra,hks}  to the distinctive pattern of the 
observed scaling violation~\cite{h1diffr} and 
 detailed features of the associated jet 
distributions.~\cite{h1diffjet,jhep02jet,klasenk}
More generally,  it suggests that the expansion  used is better justified for 
processes directly coupled to the gluon distribution than for $F_2$,   
see e.g. applications to $F_L$ (or its diffractive component~\cite{golec07}) and 
 jet final states.  
A critical  discussion,  including the quark case,  is given in Ref.~\cite{hs07}.   

Note that the question  of  how  to perform 
QCD calculations that  incorporate  multiple 
scatterings along  with perturbative evolution becomes especially compelling 
in the case of   Monte Carlo event generators~\cite{multiple-talks}.  
The  application considered above   deals  with corrections to 
$Q^2$ evolution, i.e.,   a picture based on 
 strongly ordered   $k_\perp$'s.  
But the method relies on high-energy approximations that   may  be  better   suited for 
 extension to  
 evolution with  ordering in energies, or angles. It could thus  more 
likely be adapted to 
 the modeling of multiparton processes 
in  Monte Carlo generators~\cite{gustaf,jung} based on   high-energy evolution equations.

\end{document}